\documentclass[twocolumn,showpacs,preprintnumbers,amsmath,amssymb,prl,superscriptaddress,longbibliography]{revtex4-1}
\usepackage{bbm}
\usepackage{mathrsfs}
\usepackage{graphicx}
\usepackage{dcolumn}
\usepackage{bm}
\usepackage{amsmath}
\usepackage{amsfonts}
\usepackage{color}
\usepackage[colorlinks=true,citecolor=blue,anchorcolor=cyan]{hyperref}
\usepackage{floatrow}

\begin{document}

\title{Topological bands in two-dimensional orbital-active bipartite lattices}
\author{Huan Wang}
\affiliation{State Key Laboratory of Surface Physics and Department of Physics, Fudan University, Shanghai 200433, China}
\author{Jing Wang}
\affiliation{State Key Laboratory of Surface Physics and Department of Physics, Fudan University, Shanghai 200433, China}
\affiliation{Institute for Nanoelectronic Devices and Quantum Computing, Fudan University, Shanghai 200433, China}

\begin{abstract}
The search for large gap quantum spin Hall (QSH) and quantum anomalous Hall (QAH) insulators is important both for fundamental and practical interests. The degenerate multi-orbitals $p_x,p_y$ in honeycomb lattice provides a paradigm for QSH state with a boosted topological gap of the first order in atomic spin-orbit coupling. By using elementary band representation, we explore the feasibility of this mechanism for QSH in general two-dimensional lattices, and find that the biparticle lattices with $C_{3v}$ or $C_{4v}$ symmetry and degenerate multi-orbitals could work. We further provide concrete tight-binding models on honeycomb, kagome and square lattices to demonstrate the desired topological physics. By introducing ferromagnetism into QSH state, we extend the mechanism to QAH state with a boosted gap. The QSH and QAH states can be achieved when Fermi level is at integer filling only for honeycomb lattice, but at certain fractional filling for other lattices. We conclude with a brief discussion on the possible material venues for such mechanism.
\end{abstract}

\date{\today}

\maketitle

\emph{Introduction}. 
The search for topological states of quantum matter has become one of the central goals in condensed matter physics. The discovery of topological insulators (TIs) in both two and three dimensions~\cite{kane2005a,kane2005b,bernevig2006d,koenig2007,fu2007,qi2008,Chen2009,zhang2009,xia2009,hasan2010,qi2011,wang2017c}, quantum anomalous Hall (QAH) effect~\cite{haldane1988,liu2008,yu2010,chang2013b,wang2015d,liu2016} and other topological states have significantly enriched the variety of quantum matter, and may lead to potential applications in electronics and quantum computation~\cite{nayak2008}. The dissipationless helical/chiral edge states in  quantum spin Hall (QSH)/QAH insulators are promising for the realization of low-power-consumption electronics, while $k_BT\ll E_g$ is needed to suppress the thermal excitation of edge carriers into the bulk electronic bands~\cite{glazman2013}, here $E_g$ denotes the bulk gap. Therefore, finding QSH/QAH materials at room temperature becomes an important task in this field.

Ever since the realization of QSH effect in HgTe quantum well~\cite{bernevig2006d,koenig2007}, enormous QSH materials have been theoretically predicted~\cite{ren2016} and an exhaustive search in material database has been done recently based on symmetry indicator and topological quantum chemistry~\cite{bradlyn2017,po2017,zhangt2019,vergniory2019,tang2019,wangd2019}. Among these predicted QSH materials, most of them fall into two general mechanisms. One is the Kane-Mele mechanism for graphene~\cite{kane2005b}, where the topological gap at Dirac point is at the  level on the second-order of atomic spin-orbit coupling (SOC) and is therefore tiny. The other is band inversion mechanism for example in HgTe quantum well~\cite{bernevig2006d}, GaSb/InAs heterostructure~\cite{liu2008a,knez2011} and 1T$'$ WTe$_2$~\cite{qian2014,wu2018}, where two orbitals with opposite parities have an inverted gap $E_{inv}$ at certain high-symmetry point in the Brillouin zone (BZ) due to atomic SOC, the topological gap opens at a finite wavevector $\delta k$ away from the band inversion point due to orbital hybridization and is small compared to atomic SOC as $E_g/E_{inv}\propto\delta ka$, here $a$ is lattice constant. The gap of these QSH materials are less than $0.2$~eV~\cite{wangd2019}. Similarly, a small topological gap for QAH insulators is also expected from the spin-polarized band inversion mechanism~\cite{liu2008,wang2015d}. 

The third mechanism for QSH insulator with a boosted topological gap is to introduce degenerate multi-orbitals such as $p_x,p_y$ into honeycomb lattice~\cite{zhang2014}, it features a orbitally enriched Dirac cone and the topological gap is determined by atomic SOC without any other small prefactor, thus can be quite large in some heavy element. This was first initiated in orbital-active ultracold atom system in honeycomb lattice~\cite{wu2007}, and then found in materials with a honeycomb lattice through substrate orbital filtering, such as monolayer bismuthene on silicon~\cite{zhou2014} and SiC substrate with a large topological gap ($0.8$~eV)~\cite{hsu2015,reis2017,li2018}, and chemically functionalized stanene~\cite{xu2013} and bismuthene~\cite{liu2014}. In this paper, we study this mechanism with a boosted QSH gap in the context of elementary band representation (EBR)~\cite{zak1980,zak1981,michel1999,michel2001,bradlyn2017,po2017,cano2018}, and explore its feasibility in other two-dimensional (2D) lattices. We find the biparticle lattices with $C_{3v}$ or $C_{4v}$ symmetry and degenerate multi-orbitals could work, and provide concrete tight-binding models on honeycomb, kagome and square lattices to demonstrate the topological physics. By introducing ferromagnetism into the QSH state, we further extend the mechanism to QAH state with a boosted gap. We conclude with a brief discussion on the possible material venues for such mechanism.

\begin{figure}[t]
\begin{center}
\includegraphics[width=3.4in,clip=true]{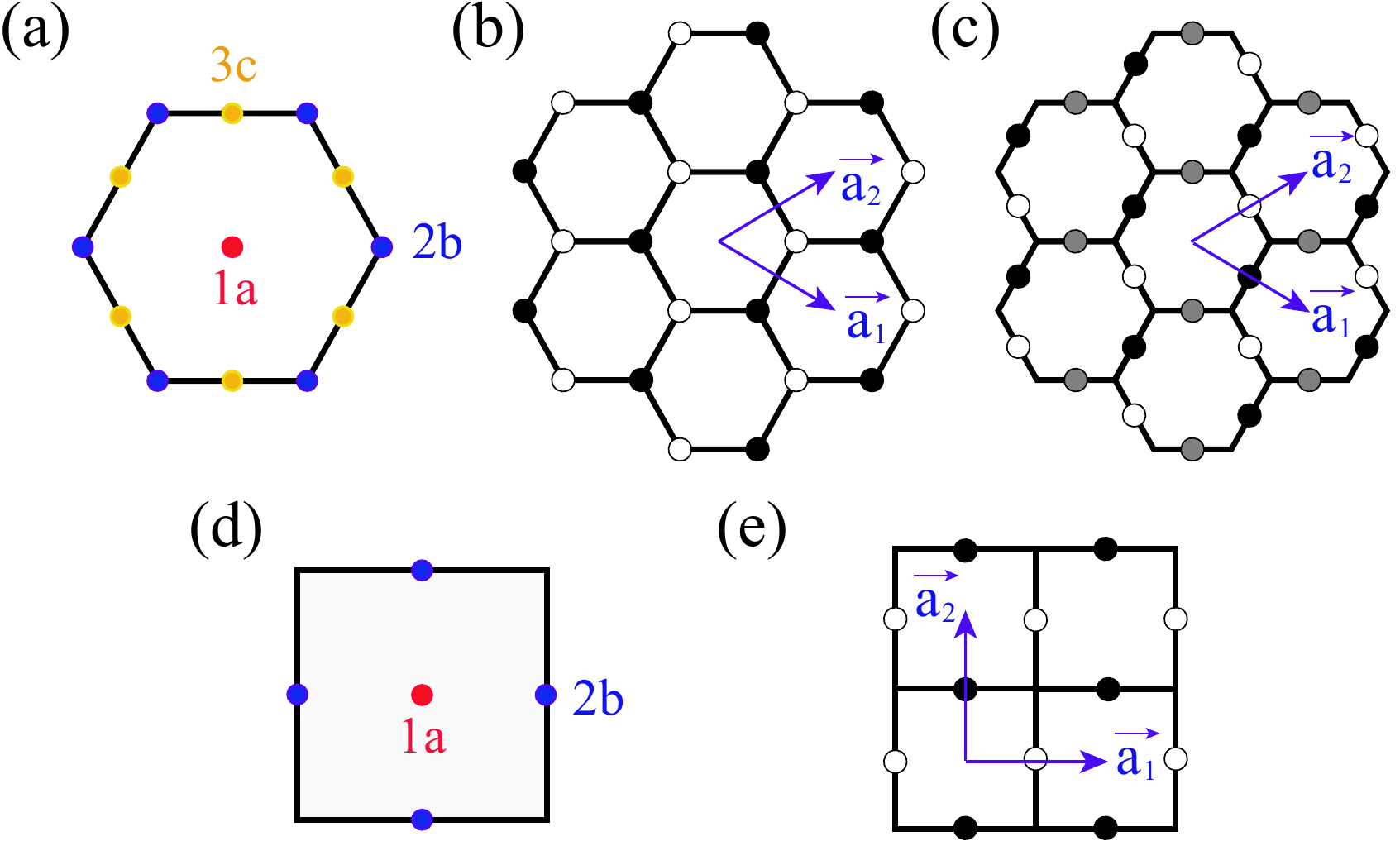}
\end{center}
\caption{(a) The maximal $1a$, $2b$ and $3c$ Wyckoff positions of the hexagonal lattice. (b) Honeycomb lattice from Wyckoff position $2b$. (c) Kagome lattice from Wyckoff position $3c$. (d) The maximal $1a$ and $2b$ Wyckoff positions of 2D tetragonal lattice. (e) Square lattice from Wyckoff position $2b$ in (d). $\vec{a}_1$ and $\vec{a}_2$ denotes the lattice vectors. The white and black dots represent A and B sublattices, respectively.}
\label{fig1}
\end{figure}

The organization of this paper is as follows. After this introductory section, next we study from symmetry analysis the ingredients for QSH state with a boosted gap. We then construct generic models for QSH and QAH and diagnose the topology from EBR. The discussion on materials and the conclusion are in the last section.

\emph{Symmetry}. Here we start from symmetry analysis to find the ingredients for QSH state with boosted gap which is on the first order of $\lambda_{\text{so}}$, here $\lambda_{\text{so}}$ is the magnitude of atomic SOC. The basic mechanism for such QSH state is to have an essential Dirac point in the absence of SOC, where the Dirac point resides at a high symmetry point in BZ and is protected by the space group and cannot be gapped without lowering such symmetry. This is equivalent to find the 2D representation in EBR for the spinless case. The rotational group $C_{3v}$, $C_{4v}$ and $C_{6v}$ satisfy such requirement. Take $C_{3v}$ as an example, the eigenvalues of the orbital angular momentum $\ell_z=0,1,-1$ mod $3$ under $C_{3v}$ are $1$, $e^{2\pi i/3}$, $e^{-2\pi i/3}$. Physically, $\ell_z=\pm1$ is related to orbitals $p_{\pm}=p_x\pm ip_y$, and time-reversal (TR) symmetry $\mathcal{T}$ transforms $\ell_z=1$ to $\ell_z=-1$ as $\mathcal{T}p_+\mathcal{T}^\dag=p_-$, which restricts them to form a two-fold degeneracy. (Here mirror symmetry $m_yp_+m^\dag_y=p_-$ also constrains to form a two-fold degeneracy.) Then we take the spin into account, instead of having the four-fold degeneracy in essential Dirac semimetal~\cite{young2015}, we expect a topological gap opened by the atomic SOC, which has the form of $\mathcal{H}_{\text{so}}=\lambda_{\text{so}}\vec{\ell}\cdot\vec{\sigma}$. This further constrains the nature of atomic orbitals. For $s$ and $p_z$ orbitals with $\ell_z=0$, $\mathcal{H}_{\text{so}}$ will not gap the Dirac point to the first order of $\lambda_{\text{so}}$. In contrast, for degenerate $p_x,p_y$ orbitals, the TR pairs $|p_+,\uparrow\rangle$ and $|p_-,\downarrow\rangle$ will split away from $|p_+,\downarrow\rangle$ and $|p_-,\uparrow\rangle$ due to $\mathcal{H}_{\text{so}}$ with finite $\ell_z=\pm1$, where the gap is on the order of $O(1)\lambda_{\text{so}}$. For graphene with a $p_z$ dominated orbital, the Kane-Mele type SOC is mediated by the on-site spin mixing which is on the order of $\lambda_{\text{so}}^2$. In the context of EBR listed in Table~\ref{table1}, there are two symmetry protected two-fold degeneracy at $\Gamma$ ($\Gamma^+_5$ and $\Gamma^-_6$) and one at $K$ ($K_6$) for spinless $p_x,p_y$ orbitals, and corresponding spinful band representation indeed is 2D but not 4D, namely the gap at Dirac cone is symmetry allowed.

\begin{table}[t]
\caption{EBRs from Wyckoff position $2b: (1/3,1/3)$ in $D_{6h}$. The irreducible representations are listed in Table I of SM, where the notation is adopted from those in the Bilbao Crystallographic Server~\cite{bilbao1,bilbao2,bilbao3,elcoro2017}.}
\begin{center}\label{table1}
\renewcommand{\arraystretch}{1.4}
\begin{tabular*}{3.4in}
{@{\extracolsep{\fill}}cccc}
    \hline
    \hline
     spinless & $\Gamma$ & $K$ & $M$ \\
    \hline
    $p_z$ & $\Gamma_{2}^- \oplus \Gamma_{3}^+$ & $K_6$ &  $M_2^-\oplus M_3^+$ \\
    \hline
    $p_xp_y$ & $\Gamma_{5}^+ \oplus \Gamma_{6}^-$ & $K_5\oplus K_1\oplus K_4$ & $M_1^+\oplus M_2^+\oplus M_3^-\oplus M_4^-$ \\
    \hline
    \hline
    spinful & $\Gamma$ & $K$ & $M$ \\
    \hline
    $p_+^\downarrow p_-^\uparrow$ & $\overline{\Gamma}_{8}^+ \oplus \overline{\Gamma}_{12}^-$ & $\overline{K}_{7}\oplus\overline{K}_{9}$ & $\overline{M}_5^+ \oplus \overline{M}_6^-$ \\
    $p_+^\uparrow p_-^\downarrow$ & $\overline{\Gamma}_{7}^+ \oplus \overline{\Gamma}_{10}^-$ & $\overline{K}_{8}\oplus \overline{K}_{9}$ & $\overline{M}_5^+ \oplus \overline{M}_6^-$ \\
    \hline
\hline
\end{tabular*}
\end{center}
\end{table}

Now we understand the two key ingredients for a boosted gap on Dirac point, one is the existence of two-fold degeneracy at high symmetry point in the absence of SOC (where all planar point group with such property are listed in SM), the other is degenerate multi-orbitals with different angular momentum $\ell_z$ such as $p_x,p_y$ and also $d$ orbitals. However, whether the boosted gap is topological or trivial is \emph{not} known without concrete models, and in general the topology of separated band cannot be diagnosed by the only high symmetry points where the boosted gap resides. Therefore, in the following we provide concrete but generic models on different 2D lattices, and diagnose band topology by EBR. 

\emph{Model}. Now we construct the tight-binding models on general 2D lattices with $C_{3v}$, $C_{4v}$ or $C_{6v}$ symmetry, which form into the hexagonal or tetragonal crystals. Since all EBRs can be induced from the irreducible representations from the maximal site symmetry group~\cite{cano2018}. Therefore, this reduce EBRs to band induced from the maximal Wyckoff positions as shown in Fig.~\ref{fig1}. Namely, EBR on the lattice related with general Wyckoff positions is the direct sum of those on maximal Wyckoff positions. The maximal Wyckoff positions on hexagonal lattice are $1a$, $2b$ and $3c$, which are related to triangle, honeycomb and kagome lattices. For diagnosis simplicity, we assume the system has inversion symmetry, where the topological $\mathcal{Z}_2$ invariant is simply determined by the parity of wave functions at the TR-invariant momentum in BZ from Fu-Kane criterion~\cite{fu2007b}. Namely, we consider $D_{6h}$ and $D_{4h}$ symmetries. We introduce the minimal degenerate atomic $p_x,p_y$ orbitals with $\ell_z\neq0$, then the generic tight-binding model is written as
\begin{equation}
\mathcal{H}_0=\sum_{\langle ij\rangle}^{a,b}c^{\dag}_{i,a}t_{ij}^{ab}c_{j,b}+\sum_{\langle\langle ij\rangle\rangle}^{a,b}c^{\dag}_{i,a}\tilde{t}_{ij}^{ab}c_{j,b}+\sum_{i,a}\lambda_{\text{so}}c^{\dag}_{i,a}\vec{\ell}\cdot\vec{\sigma}c_{i,a},
\end{equation} 
where $\langle ij\rangle$ and $\langle\langle ij\rangle\rangle$ denote the nearest-neighbor (NN) and next-nearest-neighbor (NNN) sites, respectively. $a,b=p_x,p_y$. Both NN $t^{ab}_{ij}$ and NNN $\tilde{t}^{ab}_{ij}$ transfer integrals can be simplified by symmetry considerations, and are real and spin-independent with two independent parameters. Generally $|t^{ab}_{ij}|>|\tilde{t}^{ab}_{ij}|$. The third term is atomic SOC with $\vec{\sigma}$ describes the electron spins, and in the subspace of $p_x,p_y$, $\vec{\ell}\cdot\vec{\sigma}$ reduces to ${\ell}_z{\sigma}_z$.

\begin{figure}[t]
\begin{center}
\includegraphics[width=3.4in,clip=true]{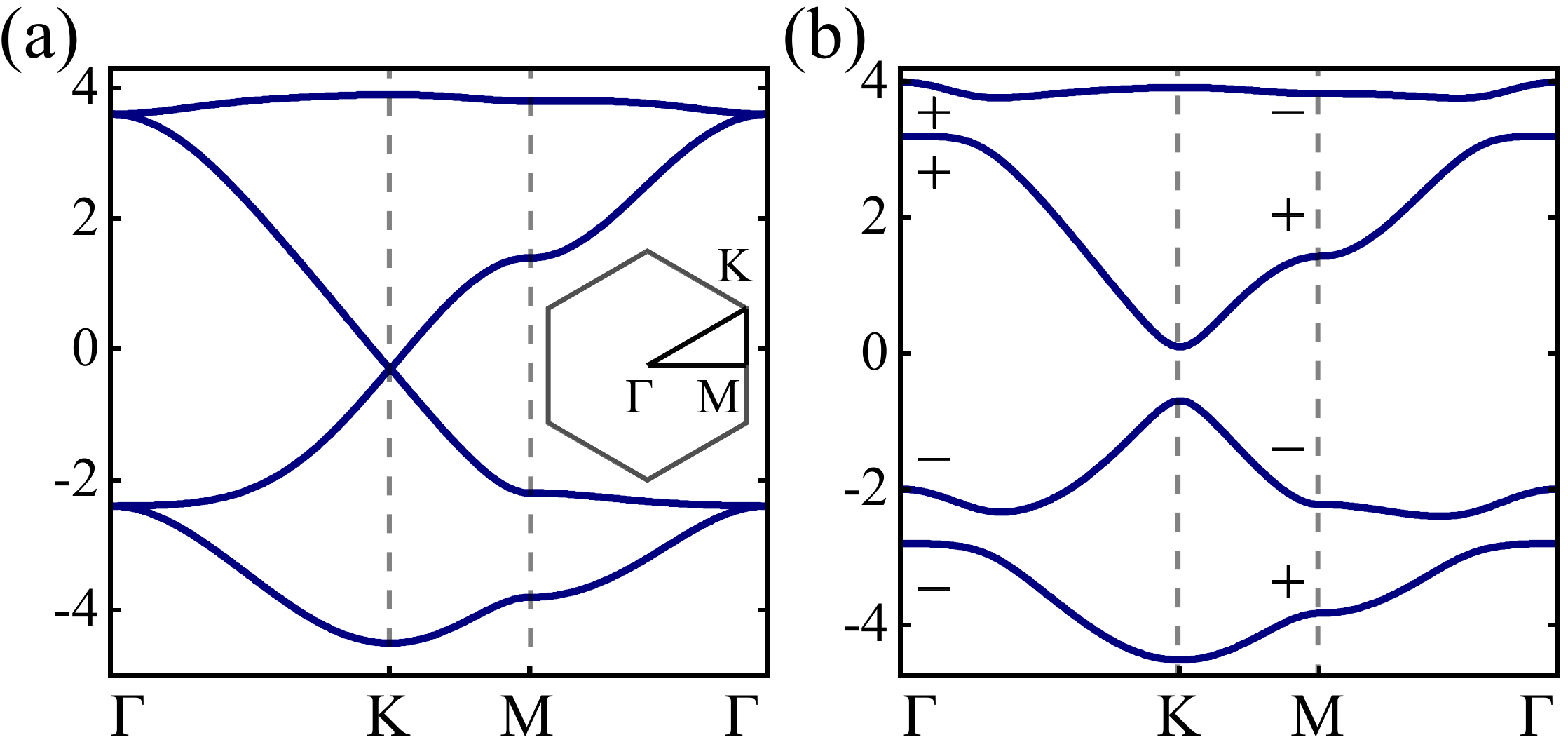}
\end{center}
\caption{(a) and (b) The band structure for $p_x,p_y$ orbitals on honeycomb lattice without and with atomic SOC, respectively. In (b), the Dirac cone is gapped by atomic SOC, where the topology of bands is labeled by the parity at $\Gamma$ and $M$.}
\label{fig2}
\end{figure}

\begin{figure}[b]
\begin{center}
\includegraphics[width=3.4in,clip=true]{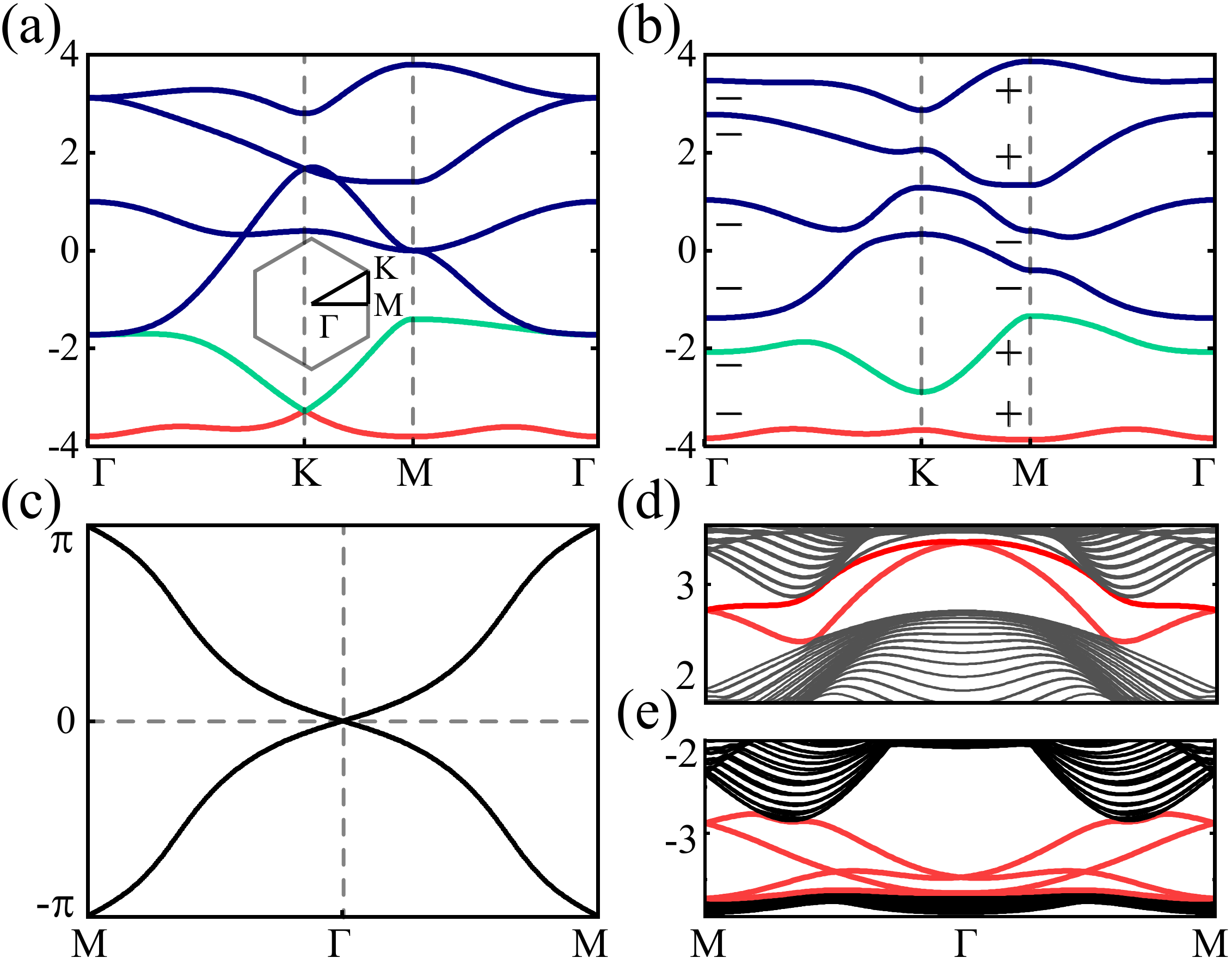}
\end{center}
\caption{(a) and (b) The band structure of $p_x,p_y$ orbitals in kagome lattice without and with atomic SOC, respectively. The parity eigenvalues are labeled at $\Gamma$ and $M$ for each bands in (b). (c) The Wilson loop for the band (red) in (b). (d) and (e) The edge spectrum with helical edge states at filling $\nu=6\mathbb{Z}\pm4$, where open boundary is along $\vec{a}_1$ direction.}
\label{fig3}
\end{figure}

\emph{QSH}. From EBR, the $1a$ Wyckoff position in $D_{6h}$ always have the same parity eigenvalue at all inversion invariant point~\cite{supplement}, namely, the parity is the same as the atomic orbitals. Therefore, the bands from $1a$ Wyckoff position cannot lead to a topological nontrivial gap (QSH) within this mechanism. Now we consider $2b$ Wyckoff position related to honeycomb lattice. The spinless spinless $p_x,p_y$ orbitals gives rise to two Dirac cones at $\Gamma$ and one Dirac cone at $K$ in Table~\ref{table1}, which is consistent with the band structure calculation shown in Fig.~\ref{fig2}(a). By adding the atomic SOC, the three Dirac cones are gapped, and the topology of each band is diagnosed by the parity eigenvalue at $\Gamma$ and $M$ shown in Fig.~\ref{fig2}(b). The QSH is achieved when the Fermi level is at filling $\nu=4\mathbb{Z}-2, 4\mathbb{Z}, 4\mathbb{Z}+2$, consistent with previous results~\cite{zhang2014,reis2017,li2018}. Similarly, the results for $3c$ Wyckoff position related to kagome lattice is shown in Fig.~\ref{fig3}. The boosted gap at $K$ is topological, which is further confirmed by parity eigenvalues and Wilson loop. Quite differently, the QSH on kagome lattice here is achieved when the Fermi level is at filling $\nu=6\mathbb{Z}\pm4$, where the edge spectrum shows the existence of helical edge states in Fig.~\ref{fig3}(d) and \ref{fig3}(e). These fractional fillings are allowed due to absence of multiple nonsymmorphic symmetries. Here we emphasis that inversion symmetry is a necessite, the Rashba SOC without inversion symmetry will not change the topology but only slightly modify the band gap~\cite{supplement}.

\begin{figure}[t]
\begin{center}
\includegraphics[width=3.4in,clip=true]{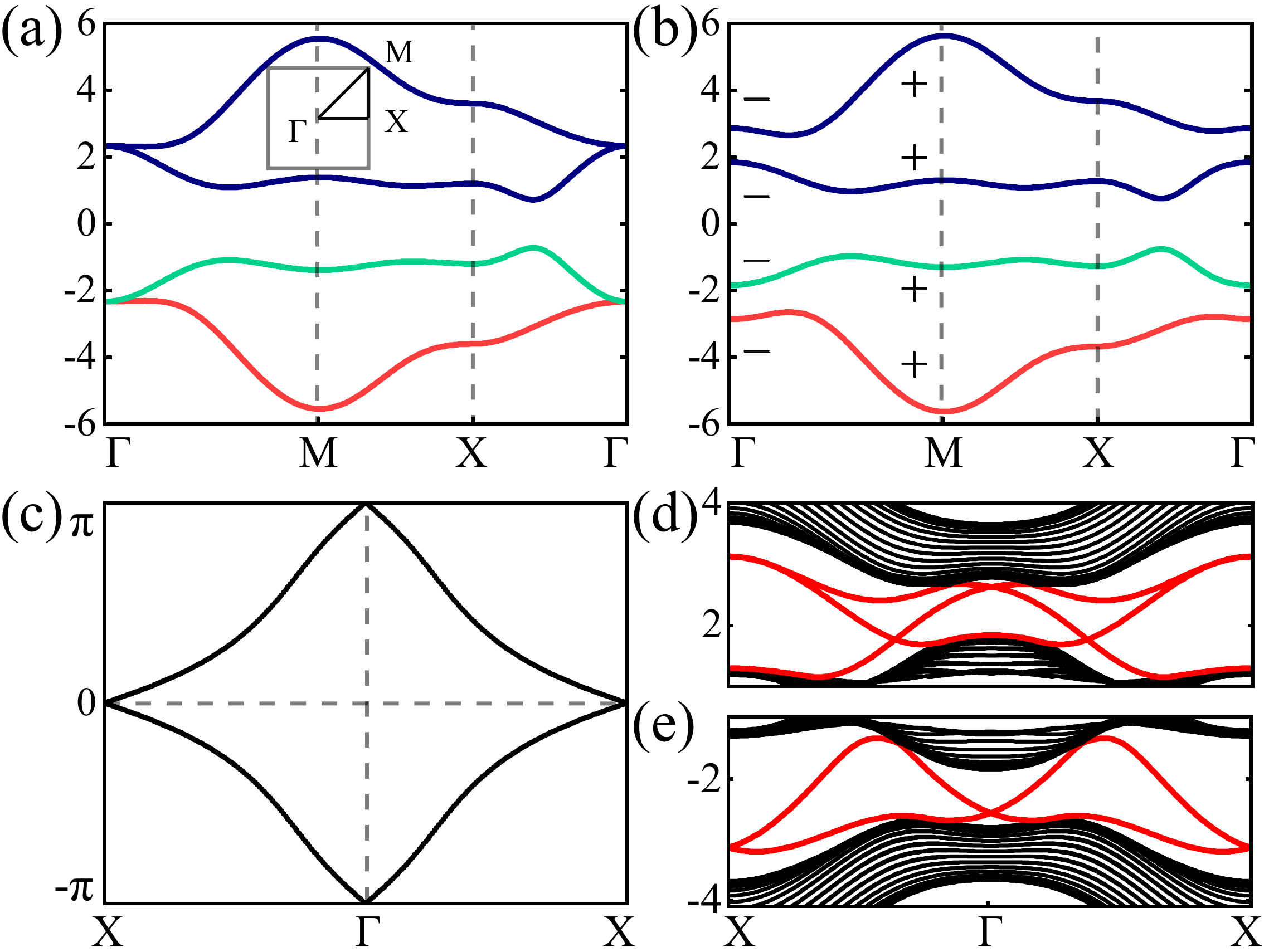}
\end{center}
\caption{(a) and (b) The band structure of $p_x,p_y$ orbitals in biparticle square lattice without and with atomic SOC, respectively. The parity eigenvalues are labeled at $\Gamma$ and $M$ for each bands in (b). Here both $X$ and $Y$ are inversion invariant momentum with the same parities, so the parity at $X$ is not labeled for it will not affect the diagnosis of topology. (c) The Wilson loop for the band (red) in (b). (d) and (e) The edge spectrum with helical edge states at filling $\nu=4\mathbb{Z}\pm2$, where open boundary is along $\vec{a}_2$ direction.}
\label{fig4}
\end{figure}

Then we study the tetragonal lattice with $D_{4h}$, there are two maximal Wyckoff positions $1a$ and $2b$ shown in Fig.~\ref{fig1}(d). Similar to the $1a$ in $D_{6h}$, here $1a$ in $D_{4h}$ cannot gives rise to QSH state by the multi-orbitals with same parity. Thus we only need to consider $2b$ Wyckoff position with the results shown in Fig.~\ref{fig4}. The QSH on biparticle square lattice is achieved when the Fermi level is at filling $\nu=4\mathbb{Z}\pm2$, and the boosted gap is at $\Gamma$. 

From the exhaustive search above, we find the 2D biparticle lattices with $C_{3v}$ or $C_{4v}$ symmetry and degenerate multi-orbitals with finite angular momentum could give rise to QSH state with a boosted topological gap of $O(1)\lambda_{\text{so}}$. It is worth mentioning that such topological gap at Dirac point on honeycomb lattice is exactly $\lambda_{\text{so}}$ without any prefactor because the crystalline symmetry will never mix different $\ell_z$, while in all other lattices the gap size is just $O(1)\lambda_{\text{so}}$ due to mixing of $\ell_z$~\cite{supplement}. Therefore honeycomb lattice is the best in view of the filling fraction and gap size.

The above conclusions on multi-orbital QSH state also apply to $d$ orbitals with $\ell_z\neq0$ if magnetism is ignored, such as $d_{xz}, d_{yz}$ and $d_{xy},d_{x^2-y^2}$ pairs. Since the choice of local orbtial is only related to the irreducible representations of local symmetry group in EBR, the above study for $p_x,p_y$ directly applies to $d$ orbitals. $d_{xz}, d_{yz}$ has effective $\ell_z=\pm1$ in both $D_{6h}$ and $D_{4h}$, while $d_{xy},d_{x^2-y^2}$ pair has effective $\ell_z=\pm2$ in $D_{6h}$ but quenched $\ell_z=0$ in $D_{4h}$. Thus the results in $d$ orbitals with finite $\ell_z$ are the same as those from $p_x,p_y$.

\emph{QAH}. The QAH effect is observed only at cryogenic temperatures owing to small energy gaps~\cite{chang2013b,mogi2015,watanabe2019,deng2020}. This is partially due to the spin polarized band inversion mechanism for QAH, where the finite wave vector away from the inversion point greatly reduces the topological gap~\cite{xu2020}. Generally, QAH state is generated from QSH state by introducing $\mathcal{T}$-breaking perturbations~\cite{wang2017c}. From the above study, we may expect a scenario where degenerate $d$ orbitals has a moderate atomic SOC but a strong spin splitting. Usually, the spin splitting for $d$ orbitals can be as large as several eV, while $\lambda_{\text{soc}}\sim0.5$~eV. This would provide a mechanism for QAH with a possible large gap.

One $\mathcal{T}$-breaking term by considering ferromagnetism is written as,
\begin{equation}
\mathcal{H}_1^m = \Delta_1\sum_{i,a}c^\dag_{i,a}\vec{\sigma}\cdot\vec{z}c_{i,a}.
\end{equation}
This term denotes the spin spitting. $\sigma_z$ is conserved, the $\uparrow$ and $\downarrow$ are decoupled. Thus this term merely shifts $\uparrow$ and $\downarrow$ bands energetically in the band structure, while the shape of the spin polarized band is unchanged as those in TR-invariant case [Fig.~\ref{fig2}(b), \ref{fig3}(b) and \ref{fig4}(b)]. The topology of the system can be simply understood from the spin Chern number. The QAH gap is maximized in the case of extreme large spin splitting where $\uparrow$ and $\downarrow$ bands are well separated. Recently, the iron-halogenide family such as FeCl$_3$ are predicted to be large gap QAH insulators, which are exactly this scenario and effectively described by $\mathcal{H}_0+\mathcal{H}_1^m$ from $|d_{xy}\uparrow\rangle, |d_{x^2-y^2}\uparrow\rangle$ orbitals~\cite{sui2020}.

\begin{figure}[t]
\begin{center}
\includegraphics[width=3.4in,clip=true]{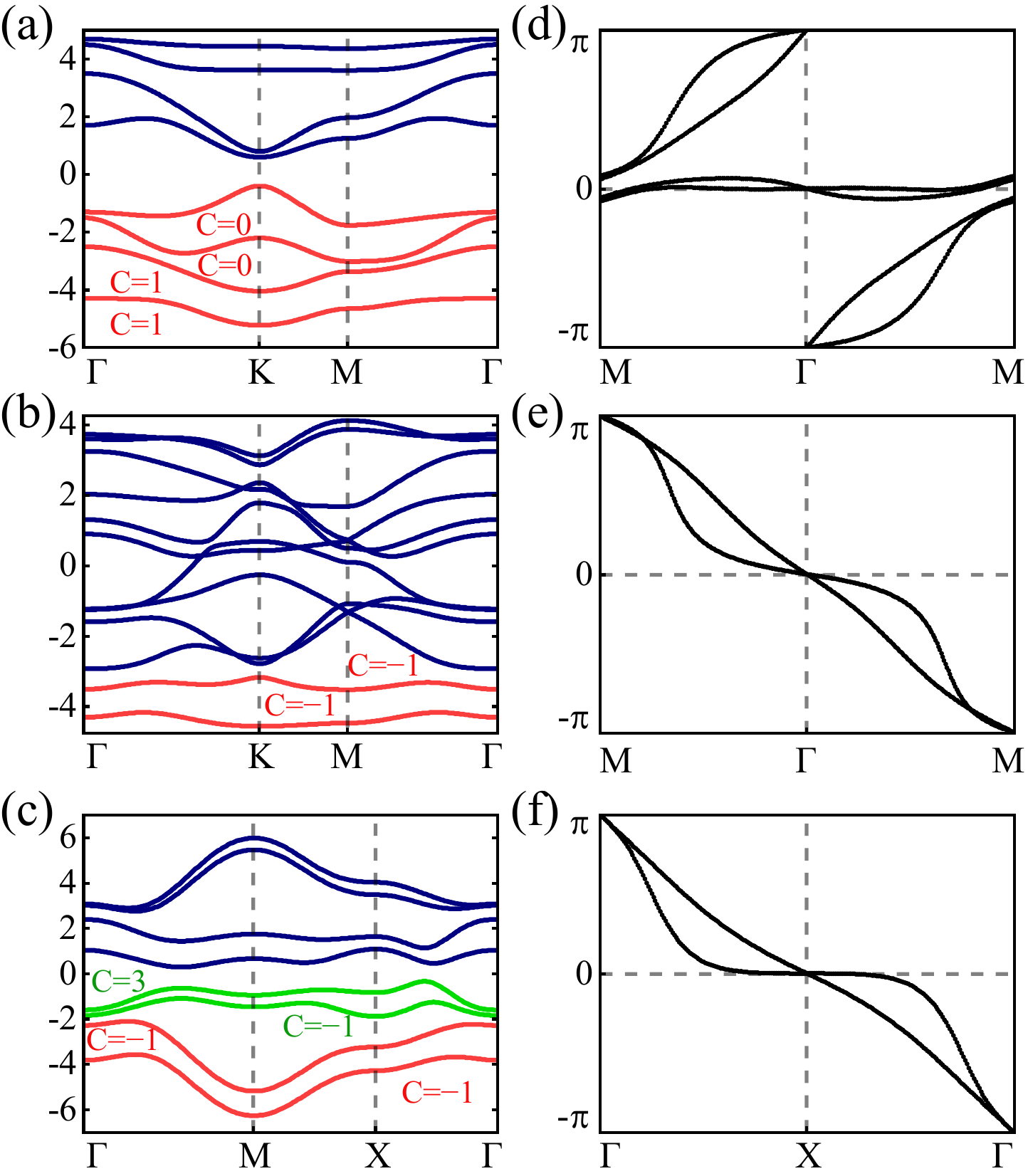}
\end{center}
\caption{ (a), (b) and (c) The band structure of the Hamiltonian $\mathcal{H}_0+\mathcal{H}_2^m$ in honeycomb, kagome and square lattices with Chern bands. The parameters are the same as those in Fig.~\ref{fig2}(b),~\ref{fig3}(b) and~\ref{fig4}(b). The corresponding Wilson loop of the bands (labeled red) are calculated in (d), (e) and (f).}
\label{fig5}
\end{figure}

Another possible $\mathcal{T}$-breaking term is
\begin{equation}
\mathcal{H}_{2}^m = \Delta_2 \sum_{i,a}c_{i,a}^\dagger \vec{j}\cdot\vec{z} c_{i,a}.
\end{equation}
This term represents Zeeman type of coupling between $p$ orbitals and localized magnetic moment, where $\vec{j}=\vec{\ell}+\vec{\sigma}$ is the total angular momentum. The band structure and topology are shown in Fig.~\ref{fig5}. When the Zeeman coupling $\Delta_2$ is competible with atomic SOC, the system becomes QAH state with Chern bands at certain filling fraction. In particular, the QAH state is achieved in honeycomb lattice when the Fermi level is at filling $\nu=4\mathbb{Z}-2$ for $\mathcal{H}_1^m$ and $\nu=4\mathbb{Z}$ for $\mathcal{H}_2^m$.

\emph{Materials.} Furthermore, we briefly discuss the search principle for realistic materials and comment on the possible candidate. From the generic models presented above, aside from appropriate crystalline symmetry, one can see the key ingredients is that the low energy physics is determined by the multi-orbitals with finite $\ell_z$. It is not energetically favorable in 2D planar materials for $p_x,p_y$ orbitals such as graphene due to $sp^2$ configuration, therefore passivation to $s$ and $p_z$ orbitals are needed. This approach is feasible by the powerful chemical functionalization of 2D materials, where the great flexibility in the chemical functional group enables forming a series of chemically new materials. Monolayer bismuthene (honeycomb lattice) on SiC substrate is a beautiful demonstration of this approach for QSH~\cite{reis2017}. By selecting 2D materials with light elements such as As/SiC~\cite{li2018}, together with Zeeman field from magnetic insulators such as MnTe, QAH state with a large gap may be realized. Recently, TbMn$_6$Sn$_6$ is found to be a kagome Chern magnet~\cite{yin2020}, where the topological physics is determined the spin polarized $d_{xy}$ and $d_{x^2-y^2}$ orbitals of Mn atom crystalline in kagome lattice and is well described by $\mathcal{H}_0+\mathcal{H}_1^m$. However, in this material the crystal field is not strong enough to push away other $d$ orbitals, resulting in a metallic state. The chemical functionalization in it with strong crystal field would make it to be insulating Chern magnet, such as in Cs$_2$LiMn$_3$F$_{12}$~\cite{xu2015}.

The multiple $d$ orbitals are quite common in transition metal oxides (TMO), which provide a large class of materials searching for topological physics. Take TMO with perovskite structure for example, the $d$ orbitals are split by the octahedral crystalline field into doublet $e_g(x^2-y^2,3z^2-r^2)$ and triplet $t_{2g} (xy,yz,zx)$ orbitals. The SOC is effective in $t_{2g}$ orbitals, and the honeycomb lattice is formed at the TMO heterostructure along (111) direction, therefore, the above study applies~\cite{xiao2011}. While the SOC is negligible in $e_g$ orbitals due to the quenched angular momentum, the strong interaction could lead to spin splitting and generate SOC dynamically~\cite{yang2011}, resulting in a QAH state described by $\mathcal{H}_0+\mathcal{H}_1^m$ in the mean field level.

\emph{Summary}. In summary, we extend the large gap QSH state on honeycomb lattice to general 2D lattices, which is based on symmetry analysis and EBR. The similar physics can happen in biparticle lattices with $C_{3v}$ or $C_{4v}$ symmetry and multi-orbitals. Here we emphasize that the global bulk topological gap is non-universal and materials specific, which should be smaller than gap at Dirac cone of order $O(1)\lambda_{\text{so}}$. However, the topological gap in orbital-active biparticle lattices is expected to larger than those from Kane-Mele and band inversion mechanism. In view of the filling, honeycomb lattice is the best for the topological physics happens at the integer filling, while other 2D lattices need fractional filling. Chemical functionalization of 2D materials provides a feasible tool to create new topological materials with these desired features, which is successful in monolayer bismuthene on SiC. Together with magnetic proximity effect, QSH state will evolve into QAH state probably with a large gap by selection of light element, where the Zeeman field competes with $\lambda_{\text{so}}$. We hope the theoretical work here can aid the search of QSH and QAH phases with large gap in new materials.

\begin{acknowledgments}
This work is supported by the National Key Research Program of China under Grant Nos.~2016YFA0300703 and 2019YFA0308404, the Natural Science Foundation of China through Grant Nos.~11774065, Science and Technology Commission of Shanghai Municipality under Grant No.~20JC1415900.
\end{acknowledgments}

\end{document}